\documentclass[manuscript,preprint]{aastex} 
\usepackage{amsmath}
\usepackage{float}
\usepackage[caption = false]{subfig}

\shorttitle{Dynamical PT in NS}
\shortauthors{Prasad et al.}

\begin{document}

\title{\bf Dynamical phase transition in neutron stars} 


\author{R Prasad}
\affil{Indian Institute of Science Education and Research Bhopal, Bhopal, India}
\email{rprasad@iiserb.ac.in} 

\and

\author{Ritam Mallick}
\affil{Indian Institute of Science Education and Research Bhopal, Bhopal, India}
\email{mallick@iiserb.ac.in}

\begin{abstract}
In this work, we have studied the dynamical evolution of the shock front in a neutron star. The shock wave is expected to possess enough strength to ignite the nuclear matter thereby converting it
to quark matter. The conversion of nuclear to quark matter is assumed to take place at the shock discontinuity. The density and pressure discontinuity is studied both spatially and temporally
as it starts near the center of the star and moves towards the surface. Polytropic equations of state which mimics real original nuclear matter and quark matter equations of state are used to 
study such dynamical phase transition.
Solving relativistic hydrodynamic equations for a spherically symmetric star we have studied the phase transition assuming a considerable density discontinuity near the center. We find that as the 
shock wave propagates outwards, its intensity decreases with time, however the shock velocity peaks up and reaches a value close to that of light. Such fast shock velocity indicates
rapid phase transition in neutron star taking place on a timescale of some tens of microseconds. Such a result is quite an interesting one, and it differs from previous calculations 
that the phase transition in 
neutron stars takes at least some tens of milliseconds. Rapid phase transition can have significant observational significance because such fast phase transition would imply 
quite strong gravitational wave signals but very short lived. Such short-lived gravitational wave signals would be accompanied with short-lived gamma-ray bursts and neutrino 
signals originating from the neutrino and gamma-ray generation from the phase transition
of nuclear matter to quark matter.
\end{abstract}


\maketitle

\section{Introduction}\label{introduction}

One of the most central topics in astrophysics is the study of compact stars (white dwarfs, neutron stars, and black holes). 
Among them, neutron stars (NSs) have proven to be excellent astrophysical laboratory to test 
the properties of matter under extreme conditions of high densities and low temperatures (see, e.g., \citep{weber,glen}).
They serve as an ideal complementary approach to the study of high-temperature 
relativistic heavy-ion collisions, which explores the high-temperature low-density regime. 

Baade and Zwicky \citep{baade} first gave the hypothesis that compact stars are made of neutrons. The verification of this conjecture was only established after the discovery of pulsar \citep{hewish} and connecting them with rotating NS \citep{gold}.
At the theoretical end, Tolman, Oppenheimer, and Volkoff were the first to solve the hydrostatic equilibrium configuration of neutron stars within the framework of general
relativity, also known as TOV equation. To close the problem and to calculate the mass and radius of NS one needs an equation of state (EoS) which describes the physical properties of matter at those densities, the relation between pressure and density, over the whole range of densities found in NSs. The search for proper EoS has motivated many researchers to study purely nuclear neutron star matter (NM).

It is known that the supernova matter is composed of relativistic neutrons, protons, electrons, and degenerate neutrinos. It is also expected that fundamental component of NS would have these relativistic neutrons, protons, electrons
and some fraction of muons in them. They can even have massive hyperons and some condensate in them. Since the discovery of the first pulsar, there has been numerous study and constant improvement while describing the matter which makes up NSs. The studies include a large number of many-body calculation
where the nucleons interact via scalar, vector, and pseudo-vector mesons. The maximum mass of such NSs lies in the range between $1.8-2.5 M_{\odot}$ \citep{li,fuchs}.
These calculations are compatible with the measured value $1.97 \pm 0.04 M_{\odot}$ for
the mass of the pulsar PSR J1614−2230 \citep{demorest,antonidis}. The core of NS can also contain other particles like
hyperons and even meson condensates. Neutron stars with a hyperon
core are sometimes referred to as “hyperon stars.” Such presence of heavy hyperons tends to make the EoS softer, thereby lowering the maximum mass, which is not compatible with new mass measurement. 
However, some recent relativistic mean-field calculations suggest that NS can support $2 M_{\odot}$ stars even after including hyperons.

The study of the evolution of NSs have revealed that 
NSs are nothing but the remnant compact stars after supernova explosion, the catastrophic gravitational collapse of massive stars (mass $M_{\odot} > 8 M_{\odot}$) \citep{haensel}, 
at the end of their evolution.
They have a radius of the order of $10$ km and a mass of around $2 M_{\odot}$ (obtained by solving the TOV equation). 
The central core of a neutron star has a density of about few times $10^{15}$ gm/cc, and the surface magnetic field ranges from $10^8 - 10^{15}$ G. At such extreme densities in the central core, the nuclear matter is no longer the stable ground state. It is prone to convert to 3-flavour quark matter (consisting of up, down, and strange quarks) which is the ground
state at such densities. The
strange quark matter conjecture \citep{itoh,bodmer,witten}  was supported by model calculations
\citep{alcock}. The most simple and popular model which describes the properties
of quark matter at such high densities is the MIT bag model \citep{chodos}. New
refined models based on results from recent experiments in laboratories have also been proposed \citep{njl,tamal1,tamal2}.
However, the quark sector is still not well understood
as the nature of strong interaction at extreme condition remains a challenge. 

Thus, the normal nuclear matter at high density and/or temperature is likely to be unstable
against QM and would eventually decay. One of the ideal laboratories for such PT is the dense core of NSs.
The problem of astrophysical phase transition (PT) is one of the main focus of this article. The PT
from normal nuclear matter to quark matter can take place at the cores of NSs where density is expected to have a value of few times nuclear density. 
Compact stars, therefore, can be made of either nuclear matter or quark matter. Stars that only contain 
nuclear matter is called NS and stars having some quantity of deconfined quark matter in them are called quark stars (QSs).
The size of the core depends on the critical density for the quark-hadron phase
transition and the EoS describing the matter phases. 

The PT from NS to QS is presumably a first-order
PT. The phase transformation is usually assumed to begin at the center of a
star where the density is at least few times the critical density. The
Several processes can trigger the PT: slowing down of a fast
rotating star \citep{glen1}, accretion of matter on the stellar surface \citep{alcock} or just cooling. Such PT could happen in newly born stars
or could occur even in old NS which is accreting matter from its companion.
It is most likely that the PT happens due to some sudden density fluctuation at the center of the star
which induces a sufficient density and pressure discontinuity. As the shock propagates outwards, it has enough energy to start a combustion process which results in the PT of NM to QM.
Recent studies have explored
different PT happening in hot proto-neutron stars \citep{bombaci,drago,mintz,gulminelli}
and cold NS with hyperons \citep{Iida}. 
The PT from NM to QM is expected to produce a major rearrangement of the NS interior and subsequently 
releases a large amount of energy as well, yielding, in principle, some detectable consequences like  gamma-ray
bursts \citep{drago,bombaci1,berezhiani,mallick-sahu}, changes in the cooling rate \citep{sedrakian},
and the gravitational wave (GW) emission \citep{lin,abdikamalov}. If detected,
all these signals could give valuable insight to constrain the
properties of matter at such extreme conditions. 

There has been a lot of studies of PT in the literature; however, the exact nature remains uncertain and controversial. In the literature one can find two very
different scenarios: (i) the PT is a slow deflagration process and never a detonation \citep{drago1,ritam-new} and (ii) the PT from confined to
deconfined matter is a fast detonation-like process, which lasts about $1$ ms \citep{bhat1,bhat2,igor}. If the process is quick burning and very violent
(detonation), there could be compelling GW signals coming from them which could be detected at least in the second or
third generation of VIRGO and LIGO GW detectors \citep{abdikamalov,lin}. The earliest calculation \citep{olinto} assumed the conversion to proceed via a slow combustion, where the conversion process depends strongly on the temperature of the star. Later studies about the stability of the conversion process \citep{horvath}, and found that under the influence of gravity the conversion process becomes unstable and the slow combustion can become a fast detonation.
The relativistic calculation was done \citep{cho} to determine the nature of the conversion process, employing the energy-momentum conservation and baryon number conservation
(also known as the Rankine-Hugoniot condition). In most of this calculation, the dynamical evolution of the shock wave was not considered, and the combustion process was categorized by employing the energy-momentum and baryon number conservation across the shock discontinuity.
However, there is still no consensus about the nature of the conversion process.

Very simplistic studies of dynamics of the phase transition are carried out first by Lin et al. \citep{lin}. They assumed that the PT occurs instantaneously
at the star core which gets converted. The outer matter remains in the nuclear phase; a mixed phase region separates the inner and outer surface. The quark matter core has a 
radius of about $5$ km. The typical time scale for the conversion to happen is of the order of $0.5$ ms assuming that the sound speed in the star quark core is about $0.3 -0.5 c$.
They also carried out a simplistic calculation of the GW emission due to such a PT induced collapse. Employing such model in Newtonian hydrodynamics, they predicted that
the gravitational strain $h$ is of the order of $3-15 \times 10^{-23}$ for a source at a distance of $10$ Mpc. The energy carried by such GW is in the range 
$0.3 - 2.8 \times 10^{51}$ ergs.
Abdikamalov et al. \citep{abdikamalov} improved similar calculation for an axis-symmetric star taking GR into account. They also introduced finite-time scale for the PT instead of 
instantaneous PT.
The GW strain $h$ for their calculation lies in the range $0.1 - 2.4 \times 10^{-23}$ for different models considered. The energy output of the GW during the first $50 ms$ of the evolution is in the range between $10^{48}- 10^{50}$ ergs. They inferred that the detection possibility of such GW signals is quite low for first-generation or in that 
sense 2nd generation detectors, but entirely possible for third generation detectors, especially for the GW signals coming from Virgo cluster.

Herzog and Ropke \citep{herzog} did another calculation of 3-D hydrodynamic simulation of the combustion of NS to QS. In their work, they used more realistic EoS both for the quark and hadronic matter instead of polytropic EoS used by previous authors. They assume that PT proceeds via slow deflagration process from the center to the surface converting NM to QM.
However, the combustion velocity or the mean propagation velocity of the flame is described by the turbulent burning speed, which results from instabilities of the flaming front. Their hydrodynamic front stops somewhere inside the star, and the outer layers are still in the hadronic phase.

In this paper, we address the dynamical PT scenario, where a shock propagation induces the PT. We study the spread of the shock from the center to the surface, and the point of the contact discontinuity which propagates with the shock is the point where the NM deconfines to QM. The shock propagation is solved using the hydrodynamic Euler equations. 
The EoS of the NM and QM is obtained by fitting piecewise polytrope to the actual EoS. The density fluctuation at the center starts the whole process, and there are pressure and
density discontinuity at the core of the star. Depending on the density of the NS ( precisely that of the nuclear matter) at any particular radius
the combusted QM density at that radius is obtained by employing the energy-momentum and baryon number conservation (look into our recent paper \citep{ritam-new} for a 
detailed discussion). The calculated QM density serves as the initial condition for the hydrodynamic differential equations. With time the shock wave propagates (the discontinuity) outwards to the periphery of the star converting NM (ahead of the shock) to QM
(behind the shock). The hydrodynamic equations govern the evolution of the shock wave along the star. 
We ensure that the point of contact discontinuity is the point where the actual PT from NM to QM is taking place. 
The large discontinuity at the center of the star ensures that the shock has enough energy to sustain the combustion process. As the shock propagates outwards to the 
low-density region, the discontinuity reduces. The PT (brought about by the shock) continues up to the point where the NM energy is higher than that of the QM density and beyond that
the shock is not able to bring about a PT. Usually, this happens at about $6-8$ km from the center depending on the EoS we choose. We thereby have a star which has a quark core and a hadronic outer layer. Such stars are known as hybrid stars.
During the conversion process, the stars suffer significant structural transformation and its contracts considerably which gives rise to gravitational waves. 

The paper is arranged in the following manner. In Section II we will be discussing our numerical techniques and the codes used to generate our results. In section III we present the equation of states (EoS) which were used to model the compact stars, while in section IV we show our results for the phase transition of NS to QS. Finally, in section V 
we summarize our findings and conclude them.

\section{Numerical methods}
In our present work, we have used the open source GR1D code\citep{Oconnor}, which is a spherically symmetric general-relativistic hydrodynamics code. In this section we summaries 
the formulation of curvature and hydrodynamics equations on which the GR1D simulation is based. After presenting the essential equations, we also describe the numerical methods 
which has been used in GR1D code. 

We have made changes to the GR1D code to suit our needs, such as to include the profiles obtained by solving TOV equation as initial configuration, to locate the position of 
shock during the time evolution, and to permit a change in the EoS in the region surpassed by shock to mimic phase transition.

\subsection{Curvature Equations}
GR1D uses spherically symmetric metric in RGPS coordinates \citep{Gourgoulhon,Romero}. The metric is given by
\begin{equation}
 ds^{2} = -\alpha(r,t)^{2} dt^{2} + X(r,t)^{2} dr^{2} + r^{2} d\theta^{2}+r^{2} sin^{2}\theta d\phi^{2}
\end{equation}
where $ \alpha(r,t)= \exp[{\Phi(r,t)}]$ with $\Phi(r,t)$ being the metric potential and $X(r,t)=\left(1-\frac{2m(r,t)}{r}\right)^{-1/2}$ with $m(r,t)$ being the gravitational 
mass at the radial distance $r$. 

Assuming the matter to be a perfect fluid, which is described by a mass current density $J^{\mu}=\rho u^{\mu} $ and stress energy tensor 
$T^{\mu \nu}= \rho h u^{\mu}u^{\nu}+g^{\mu \nu} P$
where $\rho$ is the rest mass density , $P$ is the fluid pressure, $h=1+ e +P/\rho$ is the specific enthalpy with $e$ being specific internal energy. 
$u^{\mu} = (W/\alpha,W v/X,0,0)$ is the fluid 4-velocity with $W=\sqrt{\frac{1}{1-v^{2}}}$ is the Lorentz factor and $v$ is the physical radial velocity. 

The expression for gravitational mass can be derived from Hamiltonian constraint equation and the expression for metric potential can be derived from the momentum 
constraints. It comes out to be
\begin{eqnarray}
 m(r,t)= 4 \pi \int_{0}^{r} (\rho h W^{2} -P) {r'}^{2} dr' \\ 
 \Phi(r,t) = \int_{0}^{r}  X^{2} \left[ \frac{m(r',t)}{r'^{2}} + 4\pi r' (\rho h W^{2} v^{2} + P)  \right] dr'   +  \Phi_{0} 
\end{eqnarray}
$\Phi_{0}$ is obtained by matching the $\Phi(r,t)$ at the star surface $(r=R_{*})$ to the metric potential of Schwarzschild metric given by
\begin{equation}
 \Phi(R_{*},t) = \ln[\alpha(R_{*},t)]=\frac{1}{2} \ln \left[ 1- \frac{2 m(R_{*},t)}{R_{*}}  \right]
\end{equation}

\subsection{Evolution Equations}

The GR hydrodynamics equation used in GR1D is based on flux conservative Valencia formulation \citep{Banyuls,Font,Font2} with modifications for spherically symmetric flow
\citep{Romero}.
The evolution equations for matter current density is obtained from the continuity equation 
\begin{equation}
 \nabla_{\mu}J^{\mu}=0
\end{equation}
and for the matter fields from the local conservation laws for stress energy tensor
\begin{equation}
 \nabla_{\mu}T^{\mu \nu}=0
\end{equation}

In the coordinate frame of GR1D where $ u^{\mu}=(W/ \alpha, W v/X,0,0)$ , the evolution of rest mass density is given by 
\begin{equation}
 \partial_{t}(D)+ \frac{1}{r^{2}}\partial_{r}\left(\frac{\alpha r^{2}}{X}Dv \right)=0
\end{equation}
where D is the conserved variable given by $ D= X \rho W $. 

The momentum evolution is given by

\begin{equation}
\partial_{t}(S^{r})+ \frac{1}{r^{2}} \partial_{r} \left[ \frac{\alpha r^{2}}{X}(S^{r}v+P) \right] = \alpha X \left[ (S^{r}v-\tau - D) \left(8 \pi r P+ \frac{m}{r^{2}} \right) 
+ \frac{Pm}{r^{2}} + \frac{2P}{X^{2}r} \right] + \alpha W \left( v Q_{E}^{0} +Q_{M}^{0} \right)
\label{big} 
\end{equation}
where $S^{r}$ is the conserved variable given by $S^{r}=\rho h W^{2} v$ and conserved variable $ \tau = \rho h W^{2} -P -D$. 

The energy evolution equation is given by
\begin{equation}
 \partial_{t} (\tau)+ \frac{1}{r^{2}} \partial_{r} \left[ \frac{\alpha r^{2}}{X}(S^{r}-vD) \right] = \alpha W \left(Q_{E}^{0} + v Q_{M}^{0} \right)
\end{equation}
where $Q_{E}^{0}$ and $Q_{M}^{0}$ denotes the energy and momentum source terms.
The conserved variables are function of primitive variables $\rho$, $e$ , $v$ and $P$.

This set of evolution equations can be written as
\begin{equation}
\partial_{t} \vec{U} + \frac{1}{r^{2}} \partial_{r} \left[ \frac{ \alpha r^{2}}{X}\vec{F}\right]=\vec{S}
\end{equation}
where $\vec{U}$ is the set of conserved variables, $\vec{F}$  is their flux vector and $\vec{S}$ vector contains gravitational and geometric sources. They can expressed as
\begin{eqnarray}
& \vec{U}  =\left[ D, S^{r}, \tau \right] \\ 
& \vec{F} =\left[ Dv, S^{r}v+P, S^{r}-Dv \right] \\ 
& \vec{S} =[0, \alpha X (S^{r}v-\tau - D) \left(8 \pi r P+ \frac{m}{r^{2}}\right)  +  
\alpha P X \frac{m}{r^{2}} + \frac{2 \alpha P}{Xr}  + \\
&\alpha W \left( v Q_{E}^{0} +Q_{M}^{0} \right), \alpha W \left( Q_{E}^{0} + v Q_{M}^{0} \right) ] \nonumber
\end{eqnarray}

\subsection{Numerical Method}

To develop the numerical code we first discretized the space. Then applying the method of lines (MoL \citep{Hyman}), and by using the standard second order or third order Runge-Kutta  
time integration of conserved variables is carried out.
The spatial discretization is done by finite volume approach \citep{Romero,Font2}. The variables are defined at cell centers $i$ and are interpolated at cell interfaces. 
At the cell interfaces, the inter-cell fluxes are evaluated. To interpolate, the third-order piecewise -parabolic method (PPM \citep{Colella})is used to smoothen parts of flow.  
The primitive variables are interpolated and then corresponding conserved variables are found at the cell interfaces. Also, piecewise-constant reconstruction is used, 
and to avoid oscillations near origin in the innermost three to five zones specifically the piecewise linear rebuilding \citep{Leer} is used.

After the conserved variables are evaluated at the cell interfaces, physical interface fluxes $\vec{F}_{i+1/2}$ are found with the HLLE Reimann solver.
The right hand side flux update term for $ \vec{U_{i}}$ comes out to be,
\begin{equation}
 RHS_{i} = -\frac{1}{r_{i}^{2} \Delta r_{i}} \left[ \frac{ \alpha_{i+1/2} r^{2}_{i+1/2}}{X_{i+1/2}} \vec{F}_{i+1/2}-\frac{ \alpha_{i-1/2} r^{2}_{i-1/2}}{X_{i-1/2}} 
\vec{F}_{i-1/2} \right]
\end{equation}

Gravitational and geometrical sources are not taken into consideration in flux evaluation but are coupled with the MoL integration.
Once the conserved variables are updated, the primitive variables needed for next time step has to be found. Since the primitive variables cannot be written 
algebraically in terms of conserved variables \citep{Font},  the primitive variables are found iteratively with an initial guess using $P_{old}$ (from previous time step) 
with help of the expressions :

\begin{eqnarray}
& v =\frac{ S^{r}}{\tau + D+ P_{old}} \\ 
& \rho =\frac{ D}{X W} \\ 
& e =\frac{\tau + D + P_{old}(1-W^{2})}{\rho W^{2}}-1 
\end{eqnarray}
where the $X$ is calculated from the conserved variables and $W$ is calculated using the value of $v$.

The equation of state is used to find the pressure (with the density obtained from above expression). This process is iterated using Newton-Raphson method till 
pressure value between successive iteration step differs by $10^{-10}$.
Thus in each time step, we find the $v$, $\rho$, $e$ at various cells (space discretized blocks) and finally the $P$ using iteration process.

\subsection{Modifications in GR1D}

We have used the GR1D code \cite{Font} as our initial milestone in solving the astrophysical problem of PT. The GR1D system is used to address the usual sod shock tube problem. We have made significant changes in the code to apply to our problem.
The sod tube problem has a left and right region separated by a discontinuity, where initially a higher density and pressure are given to left state as 
compared to the right and initial value of velocity on both sides are kept zero.
The sod problem in GR1D was by default for ideal gas EoS, and changes were made to use it for polytropic EoS. 
Changes were also made to set up the initial configuration as compared to that of sod problem. The initial density and pressure profiles along the radial 
distance from the center of the star was obtained by solving the TOV equation for a star with given central density.
A small increase in density near the center is also provided, to recreate the sudden density fluctuation. 

During the time evolution of neutron star profile in GR1D code, to locate the shock, we make use of a simple, intuitive method. Initially, at time $t=0$ the velocity at all radial distance of the configuration is kept zero. When the shock is created at the location of density discontinuity, and as it propagates towards neutron star surface, at any time $t$ the position of shock discontinuity is exhibited at the highest value in velocity versus radial distance plot. By analyzing the velocity data values at various distances for a given time $t$, we can find the highest velocity value 
in the configuration, and hence can locate the shock.

We have made changes in GR1D code to locate the shock at each time step, and also we have made changes to maintain two EoS (let's say EoS1 and EoS2), EoS1 in front of the shock and EoS2 behind the shock. As the shock propagates, the region surpassed by it, which in previous time step was governed by EoS1, now in the present time step is governed by EoS2. Hence it is like as the shock propagates EoS2 regulates the post-shock matter, and EoS1 regulates the pre-shock matter.
This mimics the phase transition which happens in real neutron star whereas shock propagates hadronic matter gets converted to quark matter.

\section{Equation of State}

NSs is assumed to contain mainly neutrons, protons, electrons with other baryons and leptons in a small fraction. The sigma, omega and rho mesons also form the constituent of NSs. 
In literature, different author's have considered different EoS depending upon the constituents particles they would like to discuss.

\subsection{Glendenning EOS}

The Glendenning parameter setting is chosen for describing the properties of matter in the hadronic form. 
The corresponding Lagrangian is given in the following form
\citep{serot86,boguta,glendenning} ($\hbar =c=1$)
\begin{eqnarray} 
 {\cal L}_H = \sum_{n} \bar{\psi}_{n}\big[\gamma_{\mu}(i\partial^{\,\mu}  - g_{\omega n}\omega^{\,\mu} - 
\frac{1}{2} g_{\rho n}\vec \tau . \vec \rho^{\,\mu})- \left( m_{n} - g_{\sigma n}\sigma \right)\big]\psi_{n} \\ \nonumber
 + \frac{1}{2}({\partial_{\,\mu} \sigma \partial^{\,\mu} \sigma - m_{\sigma}^2 \sigma^2 } )-
\frac{1}{3}b\sigma^{3}- \frac{1}{4}c\sigma^{4} - \frac{1}{4} \omega_{\mu \nu}\omega^{\,\mu \nu}+ \\ \nonumber
\frac{1}{2} m_{\omega}^2 \omega_\mu \omega^{\,\mu} 
-\frac{1}{4} \vec \rho_{\mu \nu}.\vec \rho^{\,\mu \nu} + \frac{1}{2} m_\rho^2 \vec \rho_{\mu}. \vec \rho^{\,\mu} 
+ \sum_{l} \bar{\psi}_{l}    [ i \gamma_{\mu}  \partial^{\,\mu}  - m_{l} ]\psi_{l}. 
\label{baryon-lag} 
\end{eqnarray}

The hadronic matter is assumed to contains baryons ($n$) and leptons ($l=e^{\pm},\mu^{\pm}$).
The leptons are assumed to be non-interacting, whereas the baryons interact with the scalar $\sigma$, the isoscalar-vector $\omega_\mu$ and the isovector-vector $\rho_\mu$ mesons. The fundamental properties of NM and that of finite nuclei are used to fit the adjustable parameter of the model.
We can use different fitting parameters to generate different EoS. In the present problem we have used Glendenning \cite{glendenning} parameter set to construct our star.


\subsection{Bag model EOS}

To describe the QM, we use simplistic MIT bag model \citep{chodos}. Such simple model is enough for our calculation because 
we are not concerned about the microscopic nature of the matter but instead the macroscopic properties, namely the mass, radius, pressure, etc.
The Recent calculation has shown that inclusion of the quark interaction in this model makes it possible to satisfy the present heavy pulsar mass bound. The potential for this model can be described as
\begin{equation}
 \Omega_Q=\sum_i \Omega_i +\frac{\mu^4}{108\pi^2}(1-a_4)+B
\end{equation}
where $\Omega_i$ is the potential for species $i$ and
$B$ is the bag constant. The model has only quarks and leptons.
$\mu$ signifies the baryon chemical potential and 
$a_4$ is the interaction parameter among the quarks, varied between 1 (no interaction) and 0 (full interaction). 
The density at the core of stars is such that it contains mostly $u, d$ and $s$ quarks. The masses of the $u$ and $d$ quarks are $5$ and $10$
MeV respectively and the mass of $s$ quark is taken to be $100$ MeV. We choose the values of $B^{1/4}=140$ MeV and $a_4=0.5$. 
Such numbers are quite a conservative choice.

\subsection{Polytropic EoS}
Employing such complex EoS in the hydrodynamic calculation is a herculean task. Usually, the hydrodynamic calculation is performed using simple ideal gas EoS.
However, they are not suitable for modeling NSs. For some astrophysics calculation, polytropic EoS are often used. In our estimation, we will use such polytropes.
Dealing with the ideal Fermi gas model containing only electrons or neutrons, it comes out that the degeneracy pressure $P$ follows a power-law dependence on mass density $\rho$,
given by $P = K \rho^{\gamma}$.
This relation between pressure and density is called as a polytropic equation of state. Here $K$ and $\gamma$ are constants.
In literature polytropic EoS has been used to construct the profiles of a white dwarf or neutron star with proper choice of $K$ and $\gamma$ \citep{Shapiro}. 
Though polytropic EoS is very simplistic in some situations, it can give a good description of the star or parts of the star.
To replicate a real EoS profile (obtained by using relativistic mean field calculation), it is better suited to map the EoS with more than one polytrope, with different $\gamma$.
Finding the right ones and constructing a proper, thermodynamic consistent, EoS with them is difficult and also the result will be not unique.
However, another way of constructing polytropic EoS was given by Bonazzola et al. \citep{Bonazzola}.
The Polytrope can be described as 
\begin{eqnarray}
& \epsilon (n) = m_B n + \frac{k \epsilon_0}{\gamma -1}\left(\frac{n}{n_0}\right)^{\gamma} \\ \nonumber
& p(n) = k\epsilon_0 \left(\frac{n}{n_0}\right)^{\gamma}.
\end{eqnarray}
Where $n$ is the baryon number density, $\epsilon$ the energy density and $p$ the pressure. $k$ and $\gamma$ are dimensionless parameters, $m_B$ is the mean baryon 
mass $m_B=931.2$MeV and $n_0$ and 
$\epsilon_0$ are arbitrary number and energy densities. For neutron star matter values around nuclear density are appropriate so for example 
$n_0=0.14 - 1.7$fm$^{−3}$ and $\epsilon_0=m_Bn_0$MeVfm$^{−3}$. With those values, suitable $k$ are around 0.05 and $\gamma ~2$. 

\begin{figure*}[h]
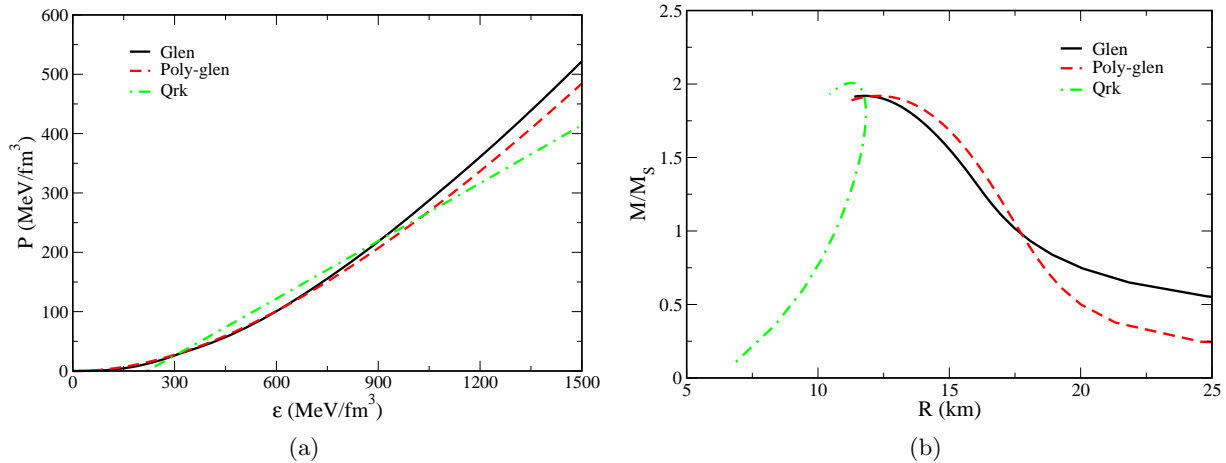

\subfloat[]{\includegraphics[width = 3.1in]{eos.eps}} \quad 
\subfloat[]{\includegraphics[width = 3.1in]{m-r.eps}}
\caption{(Color online) a) Three different EoS profiles are shown. The NM EoSs are constructed using Glendenning parameter set (bold black). 
The polytropic EoS mimicking the Glendenning EoS is shown in dashed red (represented by ploy-glen)
The QM (green-dotted) EoS is constructed using the MIT bag model with self interaction.
b) The mass-radius sequence of different compact stars (NS and QS and polytropic) are obtained by solving the TOV equations. 
The colour and marking sequence are same as that of (a).}
\end{figure*}

The EoSs are plotted in fig. 1a. We see that Glendenning and quark EoS are close to each other. All the EoS can very well be described with more than one polytrope. Fitting them with a single polytrope is a bit difficult for the whole density range, but they can be done for a specific density range which we are interested. The problem with fitting the nuclear EoS with more than one polytrope is that during the actual hydrodynamic simulation the points where we match the two polytropes give rise to substantial numerical noise. It can interfere with the real shock wave, and possibly cloud some vital physics. Therefore, in our work, we have used a single polytrope to describe nuclear and quark matter. In figure 1 we have also shown the fitted 
polytropes and we find that they match very well with the actual nuclear and quark matter EoS for a broad density range.
The polytropic EoS can generate $1.92$ solar mass NS (similar to that of 
Glendenning EoS), with $\gamma=2.12$, $n_0=0.16$ and $k=0.05$. 

In fig 1b we have plotted M-R curve for the above discussed EoS.
The Glendenning EoS generates stars with a maximum mass around $1.92 M_{\odot}$ with a radius of $12.5$ km. The fitted polytrope maximum mass and radius are quite close to that.
The stars which are more massive than about $1 M_{\odot}$ are very well represented by the fitted polytrope whereas the starts below it are ill-represented. However, in our actual 
calculation, we will use stars which are well above $1 M_{\odot}$.
The M-R sequence for the QSs generates stars with a maximum mass close to $2 M_{\odot}$ however with a much smaller radius of $10.5$ km. Such stars are composed entirely of QM 
and are not altogether stable against NM. Therefore, a pure 
QS may not be entirely stable, and its outer region may convert into NM. Such conditions can give rise to stars which have a quark core surrounded by hadronic outer 
layers known as hybrid stars (HS). 
In our calculation, we will use the quark and Glendenning EoS for the actual PT scenario as their EoS cross. Therefore, when we are considering PT, it is natural 
that NM at high density upon small fluctuation would convert to QM. 

In solving the hydrodynamic equation we have only used the polytropic ones. Solving the hydrodynamic equations with nuclear and even simple quark EoS is quite a difficult task, 
where we may have to change the whole formulation. However,
the ultimate physics does not change when we are studying the overall phase transition proceeding as a shock front and looking at it macroscopically.
We are mimicking the general nature of pressure and energy as a polytrope, and macroscopically it remains the same. The microscopic nature may also play an important role, 
however, for the moment we are more interested in the 
macroscopic behavior.

The PT is brought about by a sudden density fluctuation at the center. Somewhere near the core (for our case we are taking it to be at about $1$ km), the density fluctuation 
creates a shock like discontinuity. On the right side of the collapse, there is unburnt NM whereas, on the left side there is burnt QM. 
The shock then propagates outwards to the surface with time. We study the shock propagation by solving the hydrodynamic equations and find the spatial and temporal evolution 
of the collapse. As the shock propagates outwards, it has enough energy to make the unburnt NM burn to QM, thereby enabling a PT. The initial 
conditions of the shock are fixed from the jump conditions (energy-momentum and baryon number conservation) that was previously discussed by us \cite{ritam-new,igor}.

\section{Results}

\begin{figure*}[h]
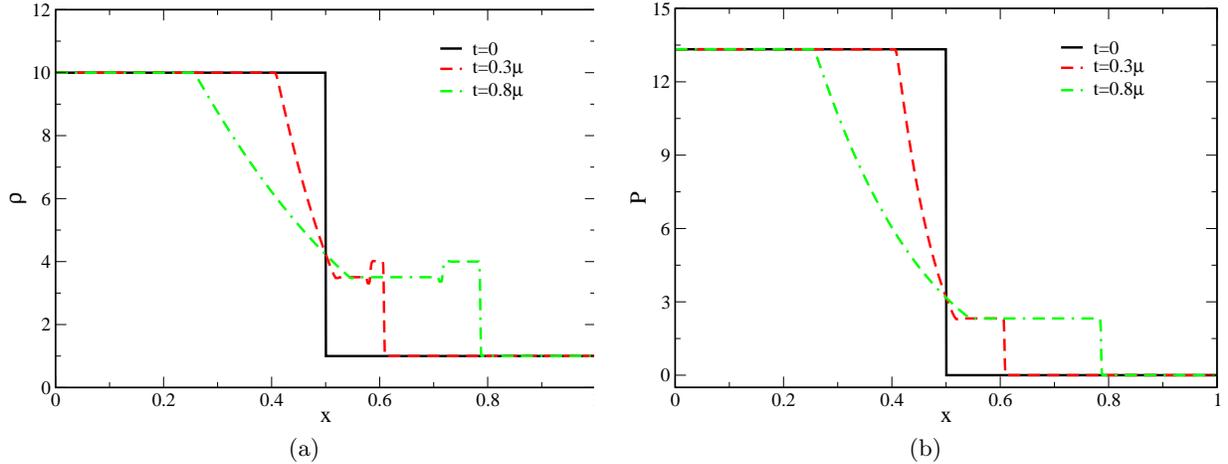

\subfloat[]{\includegraphics[width = 3.1in]{rho.eps}} \quad 
\subfloat[]{\includegraphics[width = 3.1in]{prs.eps}}
\caption{(Color online) a) Time evolution of the density is plotted as a function of $x$ for the sod problem. The evolution is plotted for two time slices $t=0.3 \mu$ and 
$t=0.8 \mu$ ($\mu$ s represents micro).  
b) The evolution of pressure as a function of $x$ for same two time slices as that of (a).}
\end{figure*}

We have modified a well established hydrodynamic GR1D code for our calculation of astrophysical PT from NS to QS. The stepping stone of this problem is the old sod shock tube problem. Therefore, as an initial check, we first show the results of the Sod shock tube problem. The problem is solved for ideal gas equation of state given by
\begin{equation}
 p=(\gamma -1)\rho e  
\end{equation}
where, $p$ is the pressure, $\gamma$ the adiabatic index and $e$ is the internal energy density. The problem is solved in the usual 1-dimension with a discontinuity separating the two sides. The discontinuity is kept at $x=0.5$, and the $\gamma$ is taken to be $1.6$. On the left side of the discontinuity, the fluid is at higher pressure and density whereas on the right side the fluid is at much lower pressure and density. With time the discontinuity propagates to the right, giving rise to shock and rarefaction waves simultaneously. This Sod shock tube problem is one of the oldest test problems for the Riemann problem where we solve the Euler's equation. The left and right side of quantities are as follows.
\begin{eqnarray}
& \rho_l=10 \hspace{0.2cm} p_l=13.2 \hspace{0.2cm} v_l=0 \\
& \rho_r=1 \hspace{0.2cm} p_r=0 \hspace{0.2cm} v_r=0.
\end{eqnarray}

\begin{figure*}[h]
\vskip 0.2in
\includegraphics[width = 3.4in]{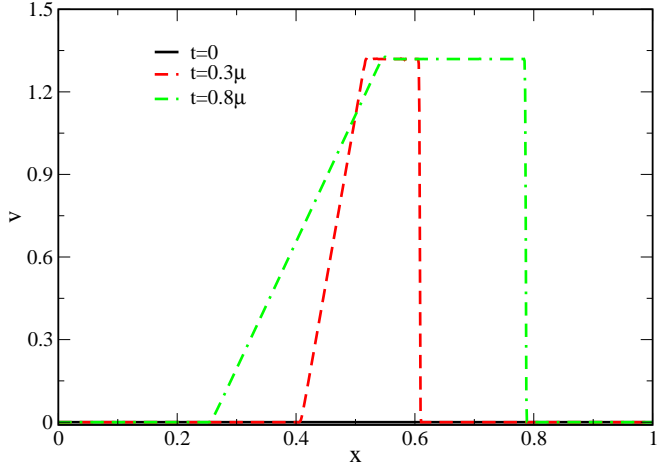}
\caption{(Color online) Time evolution of velocity as a function of $x$ is shown for two time slices ($t=0.3 \mu$ and $t=0.8 \mu$).
Initial velocity for either states (left and right) are kept at zero.}
\end{figure*}

In fig 2, we show the time evolution of the density and pressure. The evolution of the shock from the initial configuration is shown for two time slices, $t=0.3 \mu$ and $t=0.8 \mu$. As expected we find that the shock discontinuity has propagated in the right whereas the rarefaction wave has propagated to the left. 
Such behavior is found both for the density and pressure curves. 
The velocity of the shocked matter is shown in fig 3, where we find that only the shocked matter (matter affected by the shock) and rarefaction wave have attained 
some non zero velocity.
The unshocked matter velocity remains zero.
All the curves match with the usual shock tube problem.

\begin{figure*}[h]
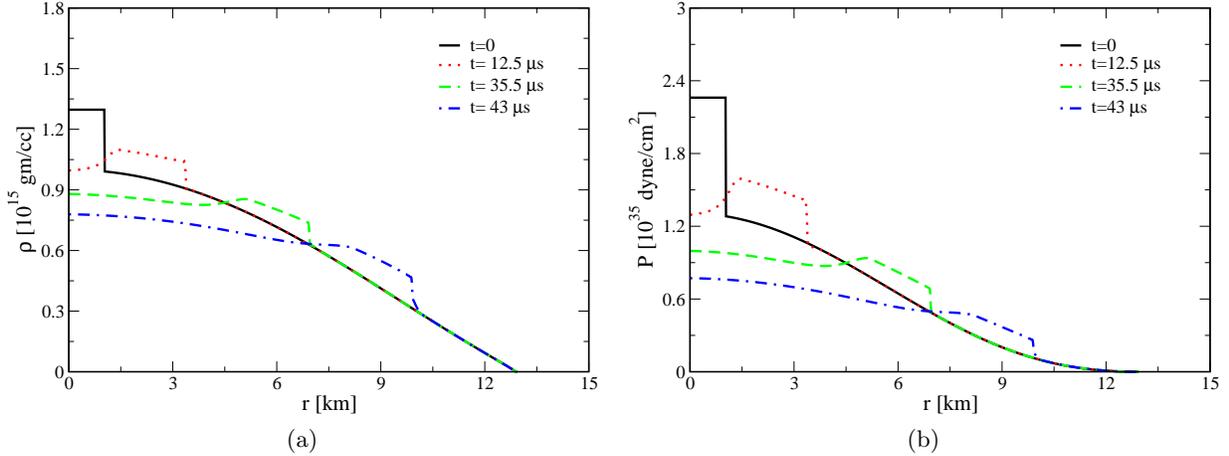

\subfloat[]{\includegraphics[width = 3.1in]{rho-s.eps}} \quad 
\subfloat[]{\includegraphics[width = 3.1in]{prs-s.eps}}
\caption{(Color online) a) Time evolution of the density as a function of radial distance from the center of the star $r$ is shown. 
The density evolution is shown for three time instances $t=0.18.6 \mu$s, $t=35.5 \mu$s and $t=43 \mu$s.
b) Pressure as a function of $r$ evolving with for three instances is shown. The time instant are same as that of fig (a).}
\end{figure*}

For the propagation of a pure shock wave in an NS, we first solve the TOV equation for a given EoS with a particular value of central density. The solution of the TOV provides 
the pressure and density as a function of the radius of the star. Using this as an input the hydrodynamic equations are solved with the given pressure and density function. 
A relativistic code is needed to solve the problem because the EoS of the NSs are usually relativistic and it also avoids the superluminal velocity problem. The problem is 
solved for a spherically symmetric star, and the discontinuity is kept near the center of the star as that is the place where the density 
is maximum. The discontinuity is not maintained precisely at the center as then it would create some numerical problem due to strong interference with the rarefaction wave. 
We should mention that we have 
assumed reflecting boundary 
condition at both ends, therefore if the initial discontinuity is kept very near the center of the star the reflecting rarefaction wave can interact with the actual shock 
wave which may hide some of the details of the shock physics. We have used a polytrope which mimics the Glendenning EoS model and have constructed a star with a central 
density of $4$ times nuclear density, which yields a star of mass $1.5 M_{\odot}$ and radius of $13$ km.
In fig 4 we plot the density (fig 4a) and pressure (fig 4b) as a function of radial distance. The density and pressure fall smoothly from the center to the surface for an 
unshocked star. We have 
kept the discontinuity at a distance of $1.05$ km from the middle of the star. We have shown $4$ time slice (at four different time including initial configuration) plots of the 
density and pressure evolution as a 
function of the radial 
distance. The initial discontinuity has to be kept a considerable amount so that we minimize the numerical fluctuation appearing due to the reflection of the rarefaction wave from 
the center. The initial discontinuity is given for the density, and the pressure discontinuity is obtained from the polytrope. The initial matter velocities at either side of the 
shock are kept to be zero. 
We see that with time the discontinuity proceeds towards the periphery of the star from the high-density to the low-density region and its strength gets reduced. Pressure plot also 
suggests similar behavior.

As the shock wave travels outwards, it carries a lot of material with it, which thereby reduces the central density of the star. Such feature is clear from the figure.
If the shock can propagate through the whole star and at the surface if it still has considerable strength, then it would expel some of the matter from the star
interior, similar to that of supernovae explosion. However, gravity should play a huge role towards the restructuring of the whole star as the shock propagates. The effect of gravity
has not been taken into account in our calculation, as it would be a very challenging task. The role of gravity and shock initial strength would play a decisive role towards restructuring of the star while the shock wave propagates.

\begin{figure*}[h]
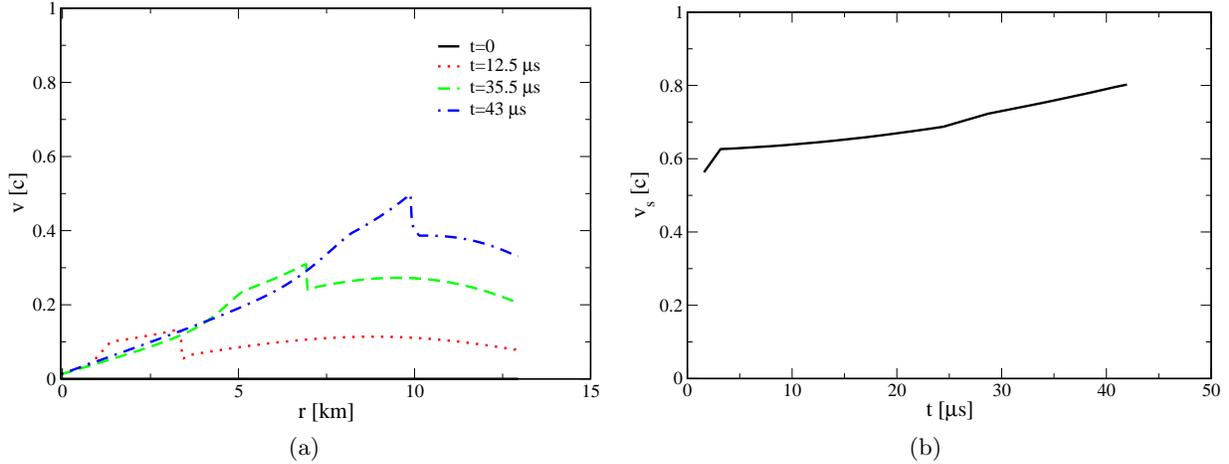

\subfloat[]{\includegraphics[width = 3.1in]{vel-s.eps}} \quad 
\subfloat[]{\includegraphics[width = 3.1in]{vs-s.eps}}
\caption{(Color online) a) Matter velocities (on either side of the shock) as a function of $r$ is shown. Their evolution is also shown for three temporal instances 
(temporal instances are same as fig 4.)
b) Shock velocity as a function of time are shown in the plot.}
\end{figure*}

In fig 5a we plot the velocity of the shocked and unshocked matter. At an initial time before the shock propagates all the matter velocities are taken to be zero. 
As the collapse spreads, the speed of both shocked and unshocked matter gets changed. With time the velocity of the matter grows as the shock propagates outward. 
The velocity profile also takes the shape of discontinuity, and the maximum speed is obtained at the shock boundary. The velocity of the unshocked matter also gets changed 
as the density profile of the star changes due to the propagation of the discontinuity (because we are dealing with a very dense closed system). The finite velocity at the shock surface indicates that the shock would not stop at the star boundary and is likely to be expelled out of the 
star. Gravity plays a decisive role in determining how much matter, if any, should be ejected from the star. The finite velocity of the shock at the 
boundary indicates that the density and pressure would also rise to a limited value at the surface. However, it is not huge. In our problem, we, therefore, have fixed the pressure to be zero at the surface.  

The location of the shock discontinuity at each slice is known from the density and pressure plots. Differentiating the shock location numerically with time we calculate the velocity of the shock. In fig 5b we plot the shock velocity with time. Initially, the shock velocity starts with $0.6$ times the light speed (c) and rises with time. The final shock velocity is about $0.8$ c. Such high shock propagation indicated that the shock could 
propagate the entire star in about $50 -100$ micro ($\mu$) seconds.

\begin{figure*}[h]
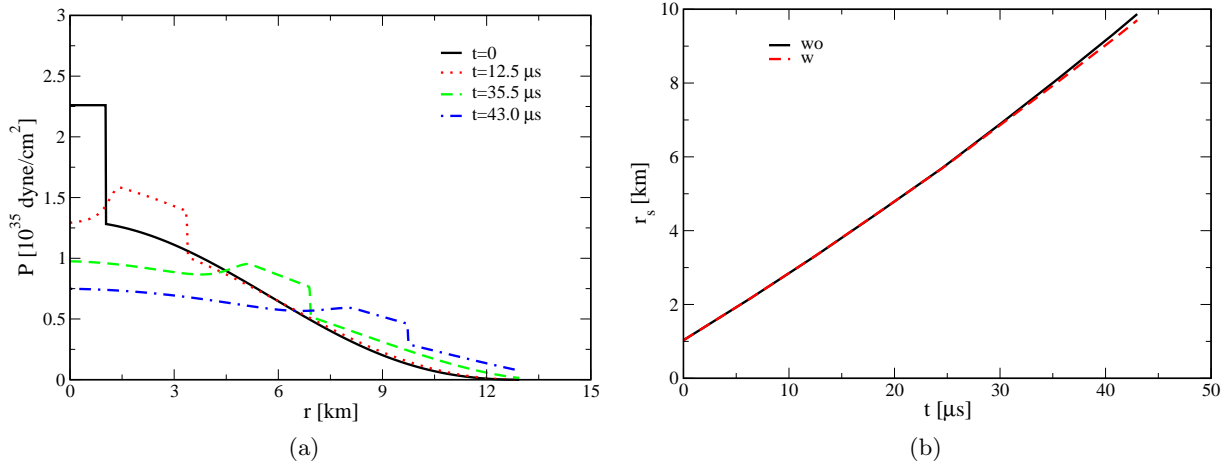

\subfloat[]{\includegraphics[width = 3.1in]{p-com.eps}} \quad 
\subfloat[]{\includegraphics[width = 3.1in]{shloc.eps}}
\caption{(Color online) a) Pressure as a function of $r$ is shown when their is no constraint on the pressure to be zero at the surface. 
Its temporal evolution is shown for three-time instances which are similar to fig 4.
b) Comparison of the shock location with time is shown with and without the constraints conditions.}
\end{figure*}

We have constrained that the pressure should be zero at the surface. However, such a strict boundary condition will not affect the shock propagation much as has been shown in fig 6. 
In fig 6a we plot the pressure evolution as the shock propagates outwards and we see that the pressure at the boundary is not zero. This is because of the boundary condition and also of the fact that we are dealing with mater at very high density. However, when we plot the shock location ($r_s$) as a function of time, we find that shock location does not change much. This also means that the shock velocity will also not differ much.

\begin{figure*}[h]
\subfloat[]{\includegraphics[width = 3.1in]{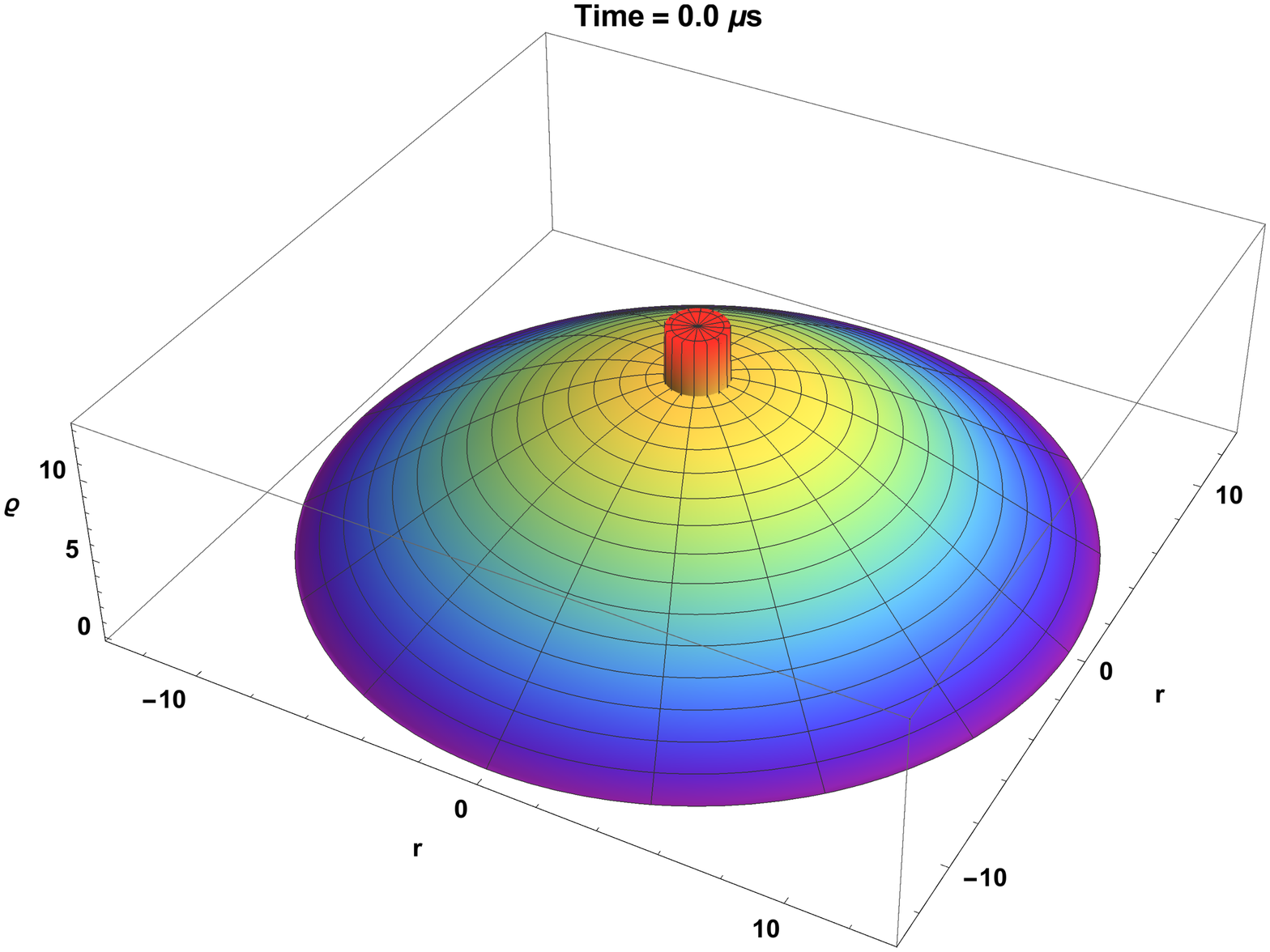}} \quad 
\subfloat[]{\includegraphics[width = 3.1in]{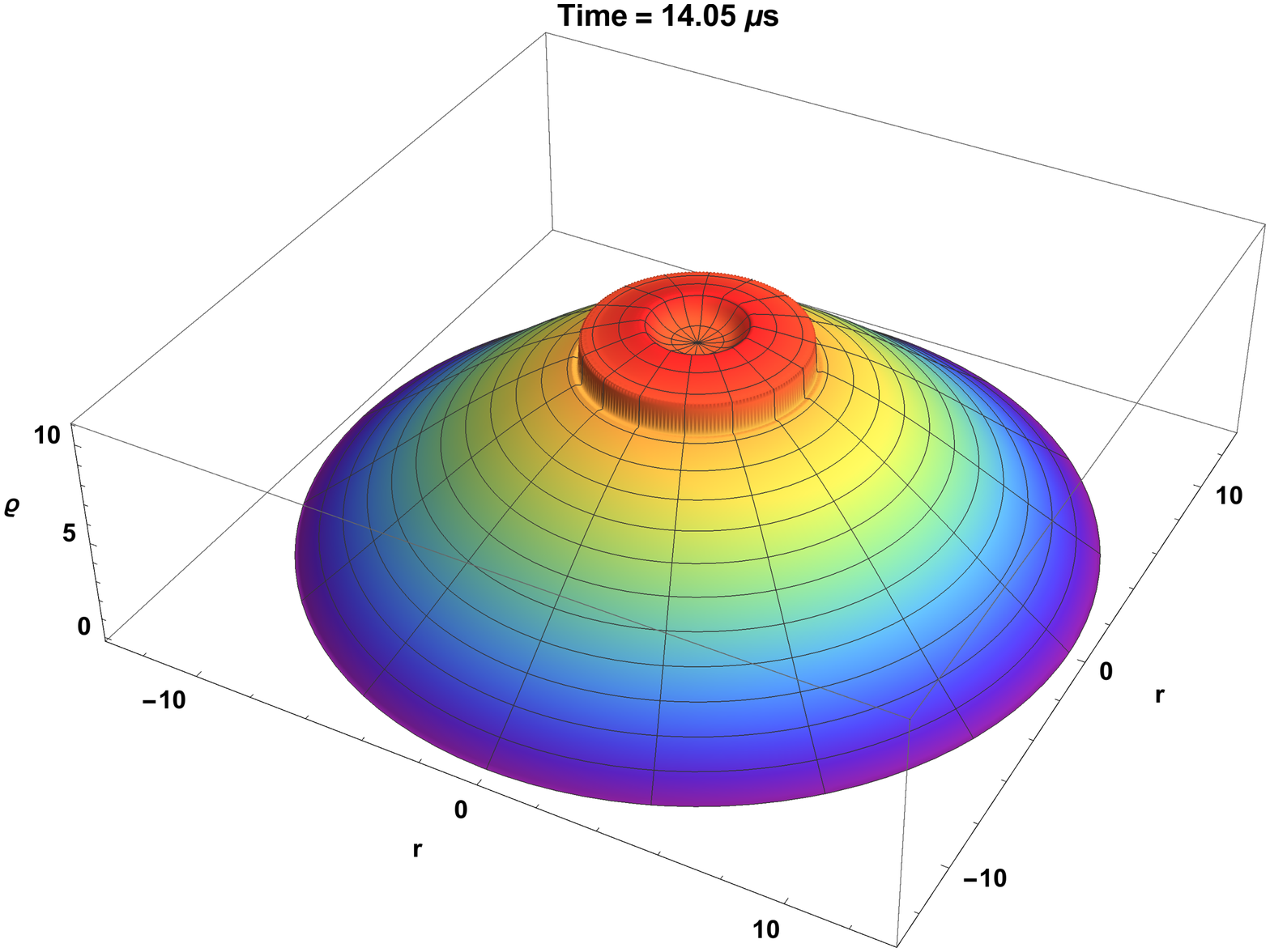}} \\
\subfloat[]{\includegraphics[width = 3.1in]{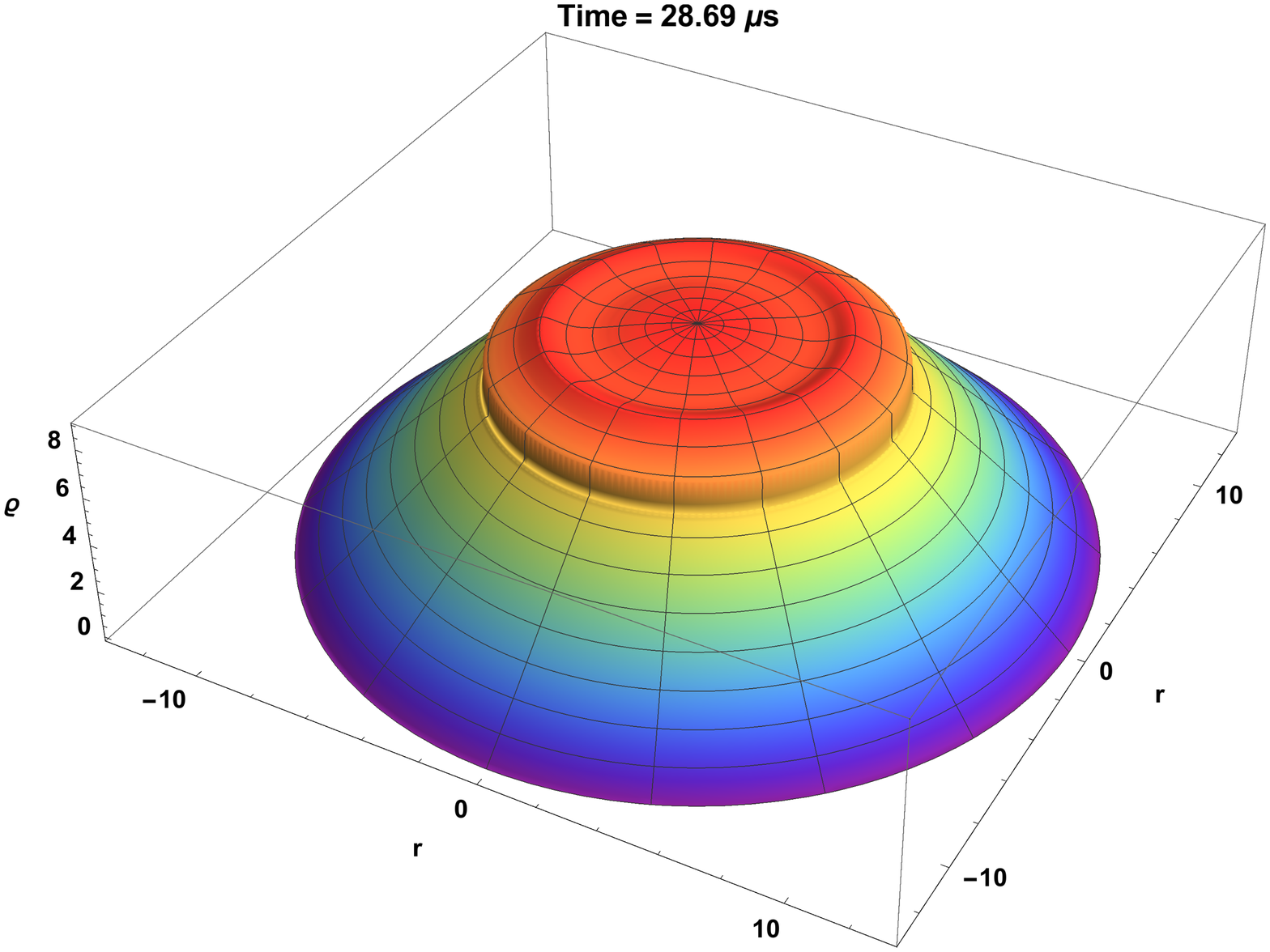}} \quad 
\subfloat[]{\includegraphics[width = 3.1in]{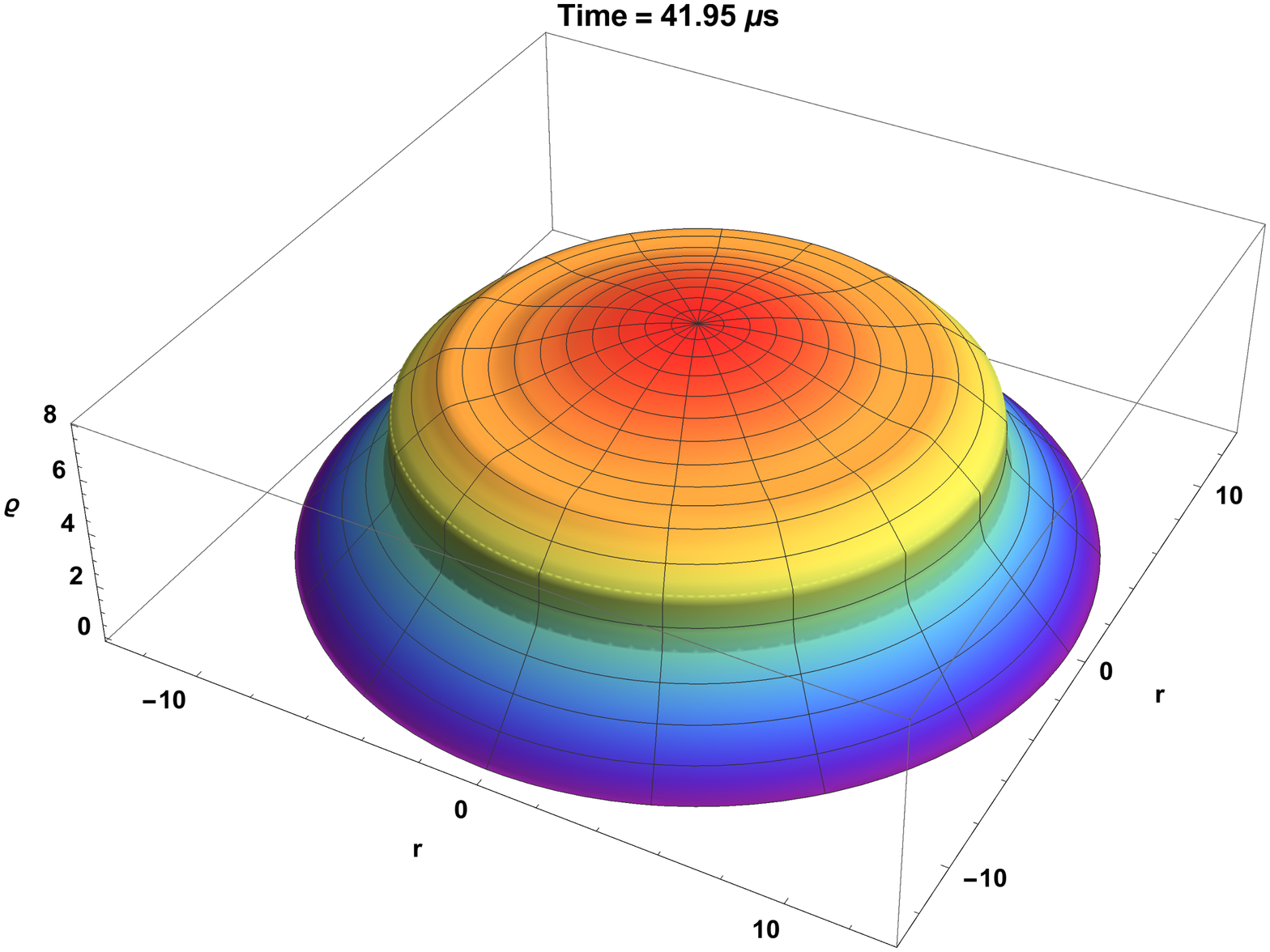}}
  \caption{Density evolution of the shock wave in the star is shown as a 2-dimensional plot. Red color signifies maximum density whereas the indigo signifies the lowest density regions.
  The evolution of the density discontinuity can be traced by following the propagation of the red shock boundary.}
\end{figure*}

The 2-d plots of the density evolution are shown in fig 7. The two-dimensional plots are obtained by rotating the 1-d plots by $360^{\circ}$ about the radial plane. 
The color coordinates show the intensity or the value of the density. The red color is for maximum
density and the purple for minimum density. As the shock propagates the discontinuity of the density follows it.

\begin{figure*}[h]
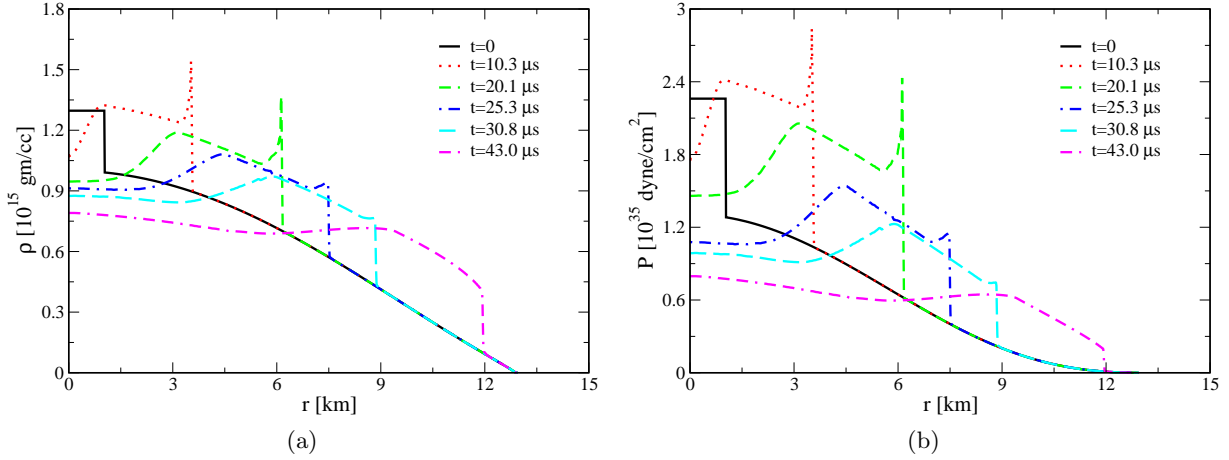

\subfloat[]{\includegraphics[width = 3.1in]{rho-2p.eps}} \quad 
\subfloat[]{\includegraphics[width = 3.1in]{prs-2p.eps}}
\caption{(Color online) a) Density as a function of $r$ is shown as it evolves in time. This density discontinuity signifies the PT from NM to QM. As the density propagates outwards
with time it means that more and more NM is being converted to QM ultimately resulting in a QS. b) Similar curve for pressure evolution is shown, where peak like pressure discontinuity
indicates the point of PT.}
\end{figure*}

Till now we have calculated the shock propagation in a star without involving PT. We currently study a scenario where a shock propagation is such that it induces a PT. 
This can be achieved numerically if we ensure that the shocked material has the properties of quark matter and the unshocked material has nuclear properties and they obey the 
energy-momentum and baryon number conservation at the shock boundary. This can be done once we locate the position of the shock discontinuity, which is done by our code. 
The code finds the shock boundary separating the two phases and then calculates the matter properties on either side of the shock, following the conservation condition. 
We assume that the shock has enough energy which can burn the unburnt matter in front of it.
We have kept the initial shock discontinuity at about $1.05$ km from the center of the star. As soon as the shock start to propagate the post-shock quantities are replaced 
by the quark EoS whereas, the pre-shock quantiles retain their values. With time the shock propagates outwards keeping such a criteria intact. Therefore, we obtain a PT from 
nuclear matter to quark matter.

In fig 8 we have shown the evolution of the density and pressure with time (displayed for six-time slices). The density and pressure of this plot differ from the previous 
growth in the fact that there is a generation of sharp peak discontinuity at the shock boundary. The peak discontinuity is present as  
the shock front propagates outwards.
The phase transition from NM to QM continues till the point where the energy of the NM is greater than that of QM. From our chosen EoS sets this happens at around $6$ km from 
the center of the star. Therefore, till $6$ km the PT is happening and the shock wave is propagating and converting the NM to QM. However, beyond $6$ km the shock wave only 
propagates through the star without converting NM to QM. From fig 8 it is also clear that beyond the $6$ km the peak suddenly disappears and there is just a small hump, which also 
dissolves as the shock wave propagates outwards. This is the unique feature of shock induced PT which has not been seen before.
We also find that the shock-induced PT is faster than the global shocks as it takes less time to reach the boundary of the star.

\begin{figure*}[h]
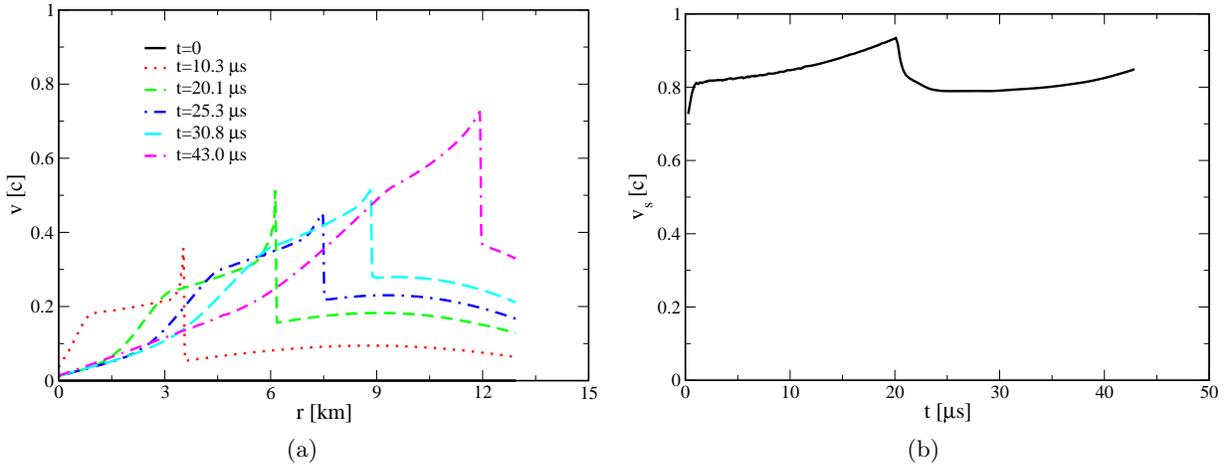

\subfloat[]{\includegraphics[width = 3.1in]{vel-2p.eps}} \quad 
\subfloat[]{\includegraphics[width = 3.1in]{vs-2p.eps}}
\caption{(Color online) a) Velocity of the QM and NM as a function of $r$ is plotted. The unshocked NM also attains some finite velocity as the shock wave is generated in the star 
although it has not reached at that point. The velocity evolution of the matter phases is shown for three time instances.
b) Shock or PT front velocity as a function of time is shown in the figure.}
\end{figure*}

Figure 9a shows the evolution of matter velocities with distance and time. As done in the earlier study we keep the matter velocity initially to be zero. At a later time, the matter velocities peak up both in the shocked and unshocked matter. 
The peaking of speed is a manifestation of the fact that we are dealing with a dense closed system and also the PT is a very fast process. 
The mater velocities for the shock-induced PT are faster than ordinary shocks. In fig 9b we plot the speed of the shock with time. The initial shock velocity
is about $0.8c$, and with time as the collapse propagates outwards the shock velocity rises and goes very close to the value of $c$ (about $0.94$c). However, at the point where the PT stops in the star and only normal shock 
propagates the velocity falls of suddenly (the peak like nature around $20\mu$ s). The velocity then grows very slowly and again reaches $0.84$ c at the star boundary.
This is due to the fact that the density of the burned matter (quark EoS) is higher than the density of the shocked but unburnt matter (nuclear EoS) for a particular radius 
(which we have obtained by energy-momentum and baryon number conservation \cite{ritam-new}. 
Such density and pressure rise in the PT scenario induces much higher discontinuity and a more violent shock. As expected the whole PT in the star is over within $50 \mu$ seconds.

In figure 10 we plot the 2-d plot of the density evolution for a shock-induced PT. The peak like feature is also visible in this scenario. The lowering of the density at the center, as the shock propagates out is also observed. As the front propagates outwards, the sharp discontinuity disappears. 

\begin{figure*}[h]
\subfloat[]{\includegraphics[width = 3.1in]{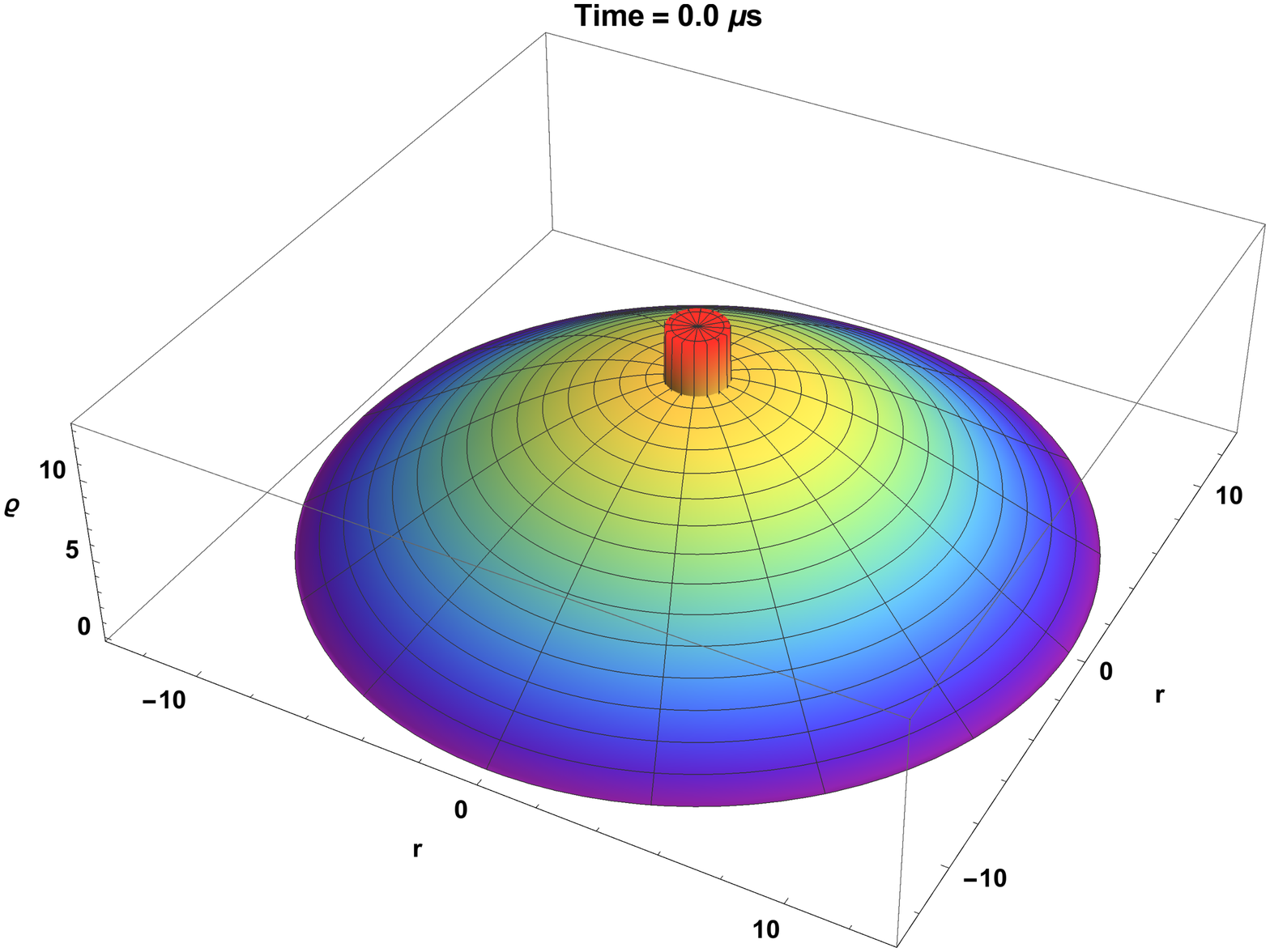}} \quad 
\subfloat[]{\includegraphics[width = 3.1in]{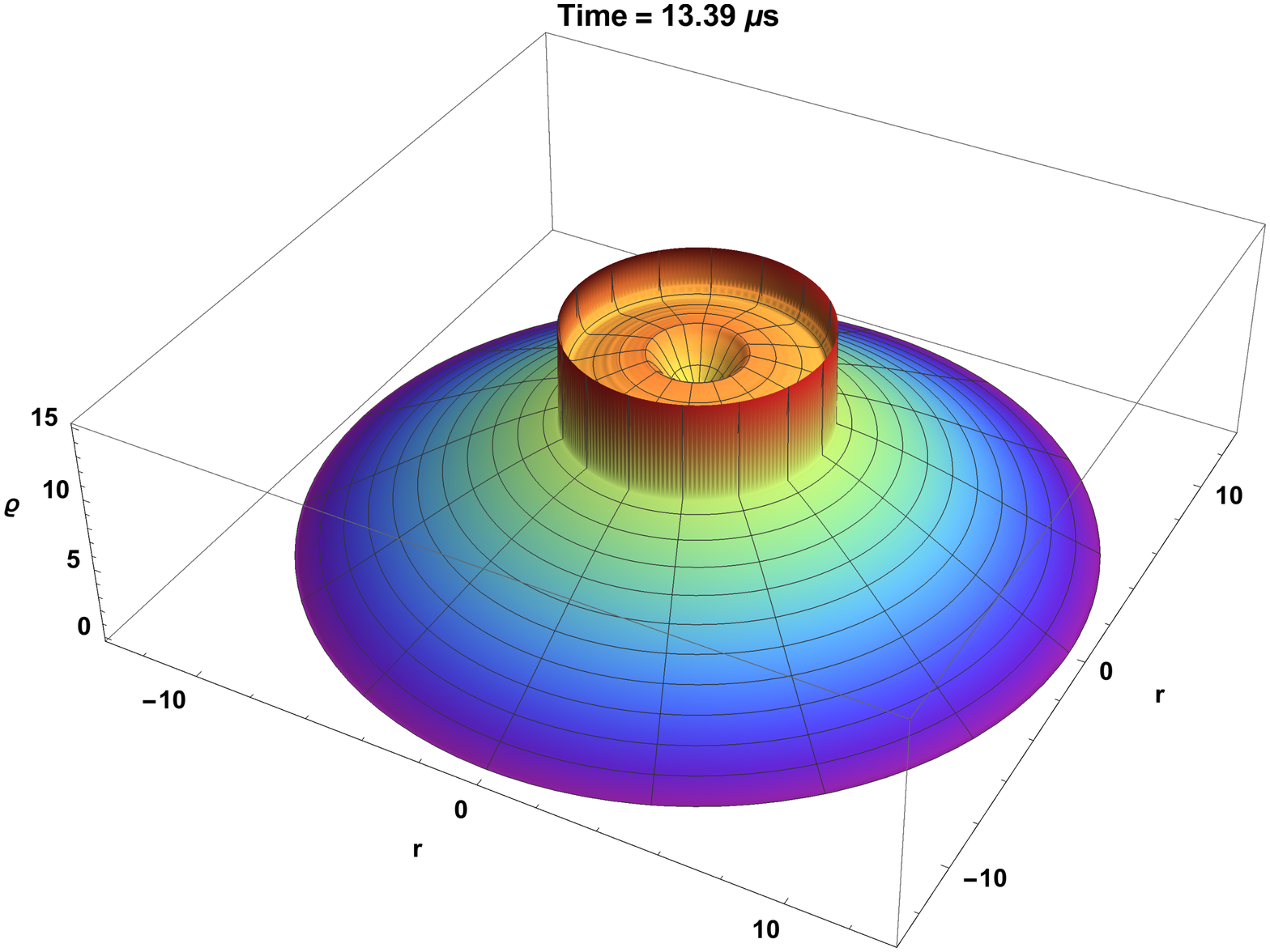}} \\
\subfloat[]{\includegraphics[width = 3.1in]{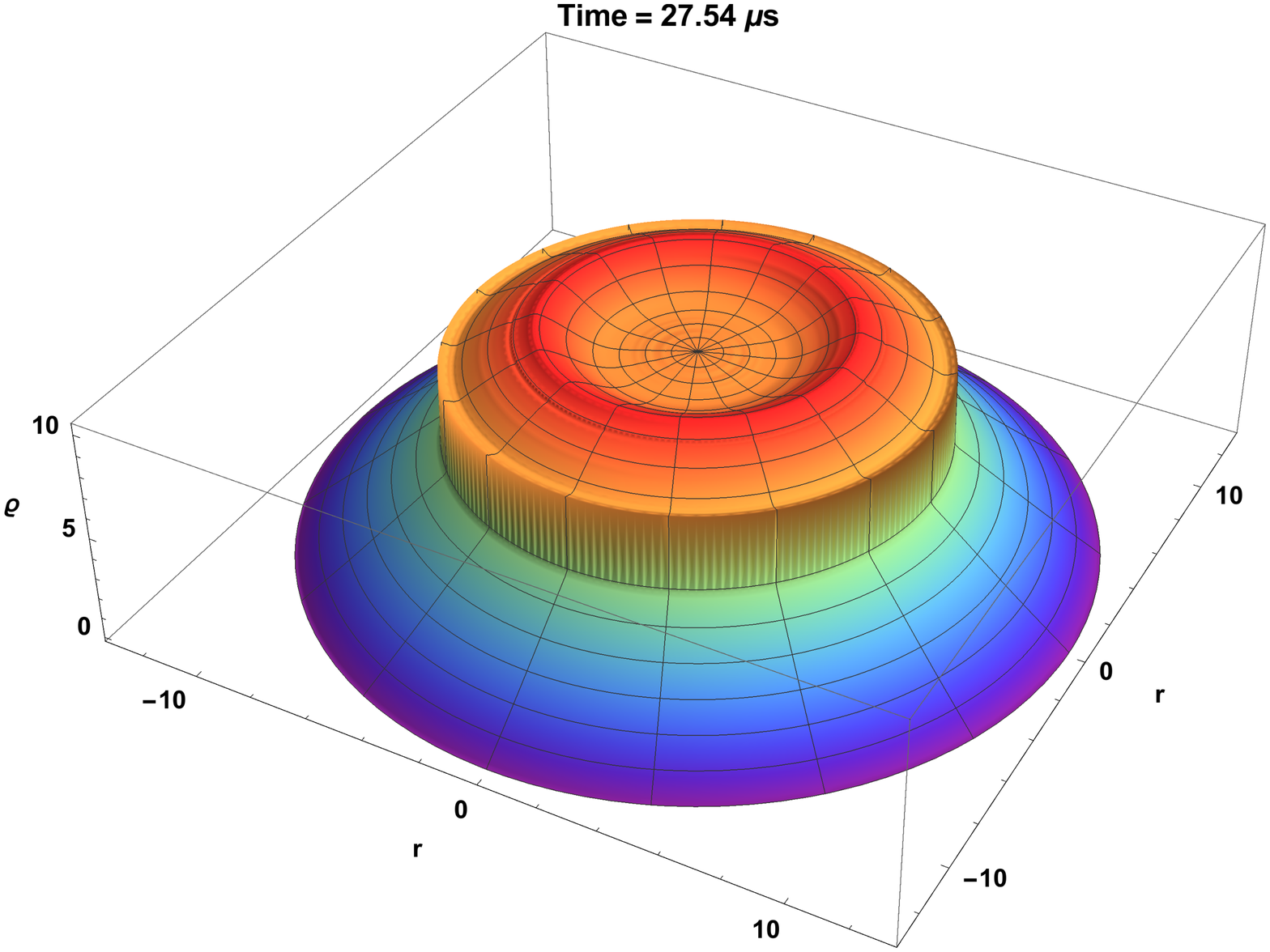}} \quad 
\subfloat[]{\includegraphics[width = 3.1in]{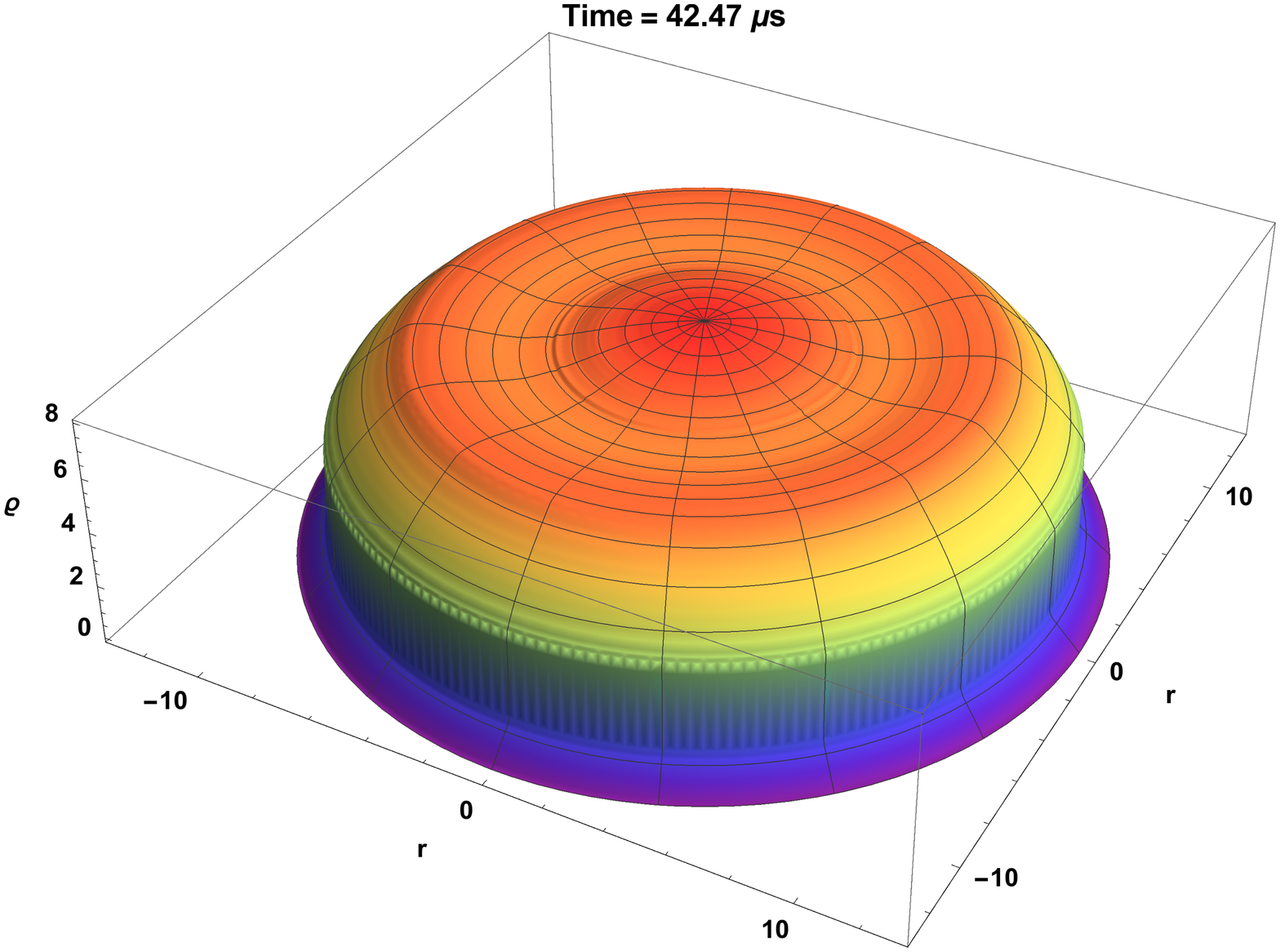}}
  \caption{2- dimensional density evolution as a function of radial distance is shown. The density evolution indicates the PT progress along the interior of the star. As the 
  density discontinuity propagates out it converts NM to QM and ultimately converting an NS to QS.}
\end{figure*}

In our calculation, we have assumed the initial matter velocity of the shock to be zero. However, it may happen that the initial speed is not zero but a finite number. However, 
at the core of the star, the initial velocity can never be very enormous, and it would not change our results much. 

\section{summary and discussion}

In this work, we have studied the dynamical evolution of the shock front as it travels from the core to the surface of a star. The shock front generates at the center of
a star due to sudden density and pressure fluctuation. The density fluctuation can be initiated by a number of reasons ranging from simple cooling to sudden braking of the star.
Once the shock front generates at such high density, it can have enough strength to ignite the nuclear matter and thereby converting it into quark matter. We have studied this 
PT from NM to QM with our hydrodynamic code. We assumed that the shock generates near the center of the star at a distance around $1$ km. In our study, we have employed 
Glendenning EoS to model the nuclear matter and simple MIT bag model with strong coupling to describe the quark matter. We have assumed that the phase transition or the burning of 
matter happens at the shock boundary and on either side of the shock, the PT happens to obey the conservation conditions. In this work, we have studied the PT using the relativistic 
hydrodynamic equations and had not taken gravity into account. 

We have modeled the PT of NS by making changes in the sod-type problem. The problem has been modified to incorporate the star profile such as the density and pressure variation as 
a function of radius (instead of ``x''). Next, we have located the position of the shock discontinuity and by our assumption, this is the place where PT 
happens. Once we have located the shock discontinuity, we have ensured that the post-shock burnt matter should have a different equation of state than the pre-shock unburnt 
matter obeying the conservation conditions. As time evolves and as the shock propagates outwards this is repeated for every small time step thereby imitating the PT scenario.

It is difficult to employ real nuclear and quark matter EoS in the hydrodynamic equations to describe matter properties. One of the ways out is to using piecewise polytropic EoS.
However, more the number of piecewise polytrope more are the chances of numerical fluctuations. To avoid such difficulty, we have employed single polytrope to describe matter 
properties. 
The single polytrope cannot precisely replicate all stars which are described by using nuclear or quark EoS. However, we made sure that the stars which would be studied for PT 
scenario are closely represented by the polytropic stars which we employ.  

In the literature, we have seen that the shock properties are affected when the matter velocities on either side of the front get very large (close to that of $c$). 
However, for the practical purpose at such high density, it is difficult for the matter to obtain such high velocity and therefore we have kept the initial speed to be zero at 
either side
of the front. We are assuming a spherically symmetric star that means our star is not rotating very fast. Studying the shock propagation we find that the matter velocity takes 
non zero values at places where the discontinuity has still not arrived. Such non-zero speed is because of the boundary condition, and also because we are dealing with dense 
confined matter. However, this does not change the pressure and density profile much, and so we have kept the surface pressure to be zero at all time. 

In the actual scenario, the real pressure may not be zero as the shock reaches the surface and thereby it may expel some matter into space. However, such a situation should be 
complemented with gravity which must be taken into account when we study the PT. Gravity is the galvanizing force which would play a huge role while restructuring of the star 
takes place after the propagation of the shock front. In our simple picture, we have not considered gravity, and we wish to address such detailed context soon.

Studying the shock propagation, we find that the shock strength decreases as the shock propagates to the boundary. However, the velocity of the shock increases at it travels outwards
to the low-density region. The shock takes about $50 - 100 \mu$ s to travel through the entire star. However, when we are studying the PT brought about by a shock we see a 
very narrow peak like structure as the shock is generated. The sharp discontinuity smooths out with time (and space). The velocity of the PT-shock is higher than regular shocks and 
almost reaches the value of $c$ as it goes to the surface. Therefore, the PT in NS over within only $50 \mu$ s. This result is a quite remarkable and different
from any other calculation done earlier. In most of the previous works, the time of PT is of the order of milliseconds. Such significant change in the shock propagation time can 
have a 
considerable significance in the determining the observational outcome of NSs, like gravitational wave and gamma-ray bursts. Such small PT time would generate a stronger 
gravitational wave, however, lasting for about $100$ micro-seconds. This is different from any other gravitational wave signals coming from other processes like black hole or 
neutron star mergers, where the 
GW last for larger times. This signal would be accompanied with other gamma-ray and neutrino signals which would come from the PT of NM to QM. The timing of the gamma-ray and 
neutrino signal would also differ from any other signal coming from another process due to the time difference. If such signals could be detected, then we can conclude that the PT
in astrophysical scenarios is a real event and there can be QS's along with NS's.

We are in the process of refining our calculation in various aspects. One is to employing real nuclear matter EoS in our problem and also employing some piecewise polytropes
and trying to reduce numerical fluctuation at the fitting ploytrope boundary. We are also trying to incorporate gravity in our problem which would galvanize the matter.
By incorporating gravity we can even have situation where as the shock propagates the radius of the star changes as there is PT from NM to QM. All such calculation are in
our immediate agenda.

\acknowledgments
RM would like to thank SERB, Govt. of India for monetary support in the form of Ramanujan Fellowship and Early Career Research Award. RM and RP would like to thank IISER Bhopal for 
providing all the research and infrastructure facilities.

\end{document}